\documentclass[pra,twocolumn,floatfix,amssymb,amsmath,secnumarabic,nofootinbib,superscriptaddress,showpacs]{revtex4-1}

\usepackage{natbib}
\usepackage{subfigure}
\usepackage{graphicx} % Include figure files
\usepackage{dcolumn}  % Align table columns on decimal point
\usepackage{bm}       % bold math
\usepackage{amsmath,amssymb}
\usepackage{psfrag}
\def\Eq#1{Eq.~(\ref{#1})}
\def\Fig#1{Fig.~\ref{#1}}
\def\co{(color online)  }

\def\la{^\lambda}
\def\A{^{\sss A}}

\def\cc{^{\rm cc}}
\def\n{n}
\def\tmu{\tilde{\mu}}
\def\sig#1#2{}

\def\bay{\begin{array}}
\def\eay{\end{array}}
\def\beit{\begin{itemize}}
\def\eit{\end{itemize}}

\def\bx{{\bf x}}

\def\e{\epsilon}

\def\F{_{\sss F}}

\def\crap#1{The following is crap:\\ #1}
\def\crap#1{\bf !!!!Erroneous entry was removed here!!!!}
\def\bei{\begin{itemize}}
\def\eei{\end{itemize}}
\def\benu{\begin{enumerate}}
\def\enu{\end{enumerate}}

\def\tv{\tilde v}

\def\sc{^{\rm sc}}

\def\s{_{\rm s}}
\def\osc{_{\rm osc}}
\def\O{{\cal O}}
\def\N{\tilde{N}}
\def\0{^{(0)}}
\def\la{^{(\lambda)}}
\def\1{^{(1)}}
\def\0{^{(0)}}

\def\2F1{_2\mathrm{F}_1}

\newcommand{\brk}[3]{\langle #1\,|\ #2 \ |\,#3 \rangle}

\def\br{{\bf r}}

\def\sig#1#2{\label{#1}\vskip -0.2cm{\small\em #1, {\bf #2}}\\}
\def\sig#1#2{\label{#1}}
\def\F{_{\sss F}}
\def\ssec#1{\subsection{#1}}

\def\barr{\begin{array}}
\def\earr{\end{array}}

\newenvironment{fanc}
{\vskip 0.5 cm\rule{1ex}{1ex}\hspace{\stretch{1}}\noindent}
{\hspace{\stretch{1}}\rule{1ex}{1ex}\vskip 0.5cm}
\def\answ#1{}
\def\answ#1{~\\ \noindent{{\bf Answer:~}\rm\color{black} #1}}

\def\eps{\epsilon}

\def\O{{\cal O}}
\def\N{{\cal N}}

\def\bea{\begin{eqnarray}}
\def\eea{\end{eqnarray}}
\def\ben{\begin{equation}}
\def\een{\end{equation}}
\def\benu{\begin{enumerate}}
\def\enu{\end{enumerate}}

\def\bei{\begin{itemize}}
\def\eei{\end{itemize}}
\def\beit{\begin{itemize}}
\def\eit{\end{itemize}}
\def\benu{\begin{enumerate}}
\def\enu{\end{enumerate}}

\def\n{n}

\def\sss{\scriptscriptstyle\rm}

\def\l{^\lambda}

\def\1var{(\bx_1...\bx\N)}

\def\half{\frac{1}{2}}

\def\br{{\bf r}}

\def\bx{{x}}

\def\s{_{\sss S}}
\def\xc{_{\sss XC}}

\def\N{_{\sss N}}
\def\H{_{\sss H}}

\def\TF{^{\rm TF}}

\def\ee{_{\rm ee}}

\def\sph_int{ {\int d^3 r}}

\def\tv{v'}
\def\la{^\lambda}
\def\A{^{\sss A}}

\def\dir{_{\sss dir}}
\def\var{_{\sss var}}

\def\cc{^{\rm cc}}
\def\sm{^{\rm sm}}
\def\osc{^{\rm osc}}
\def\co{(color online)  }

\def\inter{^{\sss int}}
\def\edge{^{\sss edge}}
\def\L{_{\sss L}}
\begin{document}
%\headheight 100pt

%%%%%%%%%%%%%%%%%%%%%%%%%%%%%%%%%%%%%%
%%             Title                %%
%%%%%%%%%%%%%%%%%%%%%%%%%%%%%%%%%%%%%%

%\sf
%\doctitle definition needed for header
\newcommand*\doctitle{Potential functionals versus density functionals}

%\draft

\title{\doctitle}

\author{Attila Cangi}
\affiliation{Max Planck Institute of Microstructure Physics, 
Weinberg 2, 06120 Halle (Saale), Germany}

\author{E.K.U. Gross}
\affiliation{Max Planck Institute of Microstructure Physics, 
Weinberg 2, 06120 Halle (Saale), Germany}

\author{Kieron Burke}
\affiliation{Department of Chemistry,
University of California, 1102 Natural Sciences 2,
Irvine, CA 92697-2025, USA}
%\affiliation{Department of Physics and Astronomy,
%University of California, 4129 Frederick Reines Hall,
%Irvine, CA 92697-4575, USA}

\date{\today}

\begin{abstract} 
Potential functional approximations are an intriguing alternative 
to density functional approximations.  The potential
functional that is dual to the Lieb density functional is defined and properties given.
The relationship between Thomas-Fermi theory as a density functional
and as a potential functional is derived.  The properties of several recent
semiclassical potential functionals are explored, especially in their 
approach to the large particle number 
and classical continuum limits.  The lack of ambiguity in the energy
density of potential functional approximations is demonstrated.
The density-density response function of the semiclassical approximation
is calculated and shown to violate a key symmetry condition.
\end{abstract}

\pacs{}

\maketitle
%\tableofcontents
%%%%%%%%%%%%%%%%%%%%
%%% INTRODUCTION %%%
%%%%%%%%%%%%%%%%%%%%
\section{Introduction and summary of results}
\label{s:intro}

Kohn-Sham (KS) density functional theory\cite{KS65}(DFT) has been an useful approach
for dealing with electronic structure problems, with more than 10,000 papers
per year being currently published.  The only approximation needed (in the
non-relativistic Born-Oppenheimer limit) is to the elusive exchange-correlation (XC)
energy  as a functional of the (spin)-densities.  While tremendous progress
has been made in constructing clever approximations\cite{B88,LYP88,B93,PBE96} over the last half century,
such approximations are generally unreliable, unsystematic, and do not produce
error estimates.\cite{RCFB09}

An alternative approach, and one that fits far better with traditional approaches
to quantum mechanics, is to consider the electronic-structure problem as a
functional of the one-body {\em potential}\cite{ES84} 
rather than of the one-body density.  However, useful approximations
beyond the local approximation\cite{T27,F28,KS65} 
are far more subtle and complex to construct,
so almost no research has been done in this area, at either the formal or the
practical level.   Notable exceptions are the pioneering work of Yang, Ayers,
and Wu,\cite{YAW04} which 
first pointed out the duality of density and potential functionals.   Thus they 
produced a deeper understanding
of the optimized effective potential.
More recently, Gross and Proetto\cite{GP09} 
emphasized the relevance of the variational principle
to PFT.
Our own recent work\cite{CLEB11} is focussed on the fundamentals 
of approximate PFT, and was motivated by recent semiclassical
potential functional approximations (PFAs) 
for simple model systems.\cite{ELCB08,CLEB10}

In the present work, we explain in detail the differences between potential and
density functionals, show that certain well-known difficulties of DFT 
are avoided, and demonstrate the accuracy achievable in PFT calculations (but
only for a model system, for which accurate PFAs have been derived\cite{ELCB08,CLEB10}).
First we give a detailed account on the exact theory and compare it with DFT.
In PFT, just as in quantum mechanics, we work within a Hilbert space
with a well-defined ground-state wave function and energy \emph{a priori} 
and therefore avoid the notoriously subtle issue of density-potential mapping  
that is required in DFT.
Another difference with DFT is that, in PFT, there are
two distinct ways of obtaining the total energy of an interacting system; 
via a direct evaluation of the functional approximations or variationally, 
through a minimization over trial potentials. We present a previously derived\cite{CLEB11} 
expression for the universal potential functional that yields 
the total energy at the variational minimum
(without having to do an actual minimization), 
if a key symmetry condition on the density-density response function is fulfilled.

Lieb\cite{L82,L83} extended the domain of the universal functional 
given in the original work of Hohenberg and Kohn\cite{HK64}. 
The construction of the Lieb density functional
involves a bifunctional of the density and potential.
What is the analog in PFT? In Sec.~\ref{s:exact.pft}, 
we repeat this exercise in PFT to explicitly show that 
the naive expression for the universal potential functional suffices. 

In practice, our results for the interacting case are not yet useful 
for an actual numerical calculation, because
they require knowledge of the interacting density
as a functional of the external one-body potential.\cite{GP09}
But all our results apply to a noninteracting system 
in some external one-body potential.
As a result we obtain an explicit expression for the noninteracting kinetic energy
as a functional of the potential.
This expression has the powerful feature that only the noninteracting density
needs to be known as a functional of the external one-body potential to fully determine 
the noninteracting kinetic energy. 

Conflating the results for the noninteracting case with the KS scheme for exchange
and correlation allows us to solve the interacting many-body problem 
in a much more efficient way than in KS-DFT, 
because there is no need to resort to the KS orbitals. 
In fact, the KS potential becomes a functional
of the external one-body potential and is determined via an alternative 
self-consistent cycle depicted in Sec.~\ref{s:app.pft}.
In principle, all this is exact; a practical
realization requires nothing but a sufficiently accurate approximation 
to the noninteracting density as a functional of the external 
one-body potential.

To illustrate how all this works and contrast it with DFT, 
in Sec.~\ref{s:TF}, we reconstruct the simplest approximation, 
Thomas-Fermi (TF) theory. 
We show how it can be constructed in either the potential functional
or density functional formalism. The logic and derivations are
completely different, but the final equations are the same.
Thus TF theory can be seen as the forerunner of exact PFT just
as easily as of exact DFT.

The early attempts of density functional construction for the kinetic\cite{K57} 
and the exchange\cite{AK85} energy begin with the local approximation
and improve upon it via a gradient expansion.\cite{HK64,KS65} 
However, those approximations fail for localized systems, 
such as atoms, molecules, or even some bulk solids.  
This is due to the presence of evanescent regions that are separated by 
turning point surfaces, i.e., where the Fermi energy cuts the potential energy surface.
In an earlier account\cite{ELCB08} we explain how this failure
can be understood by considering the expansion of the total energy
in the large-$N$ limit, where $N$ denotes the number of particles.
This analysis also explains why generalized gradient approximations 
were introduced eventually. Furthermore, our analysis suggests that 
PFAs provide a systematic approach to functional construction.
The expansion of the total energy of neutral atoms 
with respect to the atomic number\cite{S80,S81}
is probably the most prominent example of such expansions.
Considering how accurately the coefficients of such expansions 
are reproduced by an approximation also gives a measure
for the accuracy of a given approximation: we call an approximation
asymptotically exact to the $p$-th degree (AE$p$), if it yields
the first $p$ coefficients exactly.
The zeroth-order coefficient is reproduced by a local approximation, 
i.e., TF theory.
A powerful feature of such asymptotic expansions is that they yield
very accurate results even for small $N$, if the first few coefficients are
known.

PFAs beyond the simple TF approximation have been derived\cite{ELCB08,CLEB10}
for a class of simple model systems in one dimension. 
In Sec.~\ref{s:asymp} we exemplify the accuracy of PFAs by calculating the expansion
of the total energy in the large-$N$ limit for a generic, smooth one-body potential. 
Another limit in which TF theory becomes exact is the classical continuum limit, 
which we introduced as a device to derive the leading corrections 
to the TF density approximation.\cite{CLEB10}
We also assess the accuracy of existing PFAs in the classical continuum limit.
In the course of this analysis we point out another difficulty of DFT that 
is not present in PFT. Functional construction in DFT is based on approximations
to the XC energy. However this causes an intrinsic ambiguity in the energy density, 
because any term whose integral over space vanishes -- such as $\Delta f(n(\br))$ -- 
might be hidden in its definition\cite{BPE98}. This issue has been coined as the ``gauge'' ambiguity 
in energy densities for density functionals\cite{CLB98,BCL98,TSSP08}. In PFT the
energy densities are approximated directly. Therefore such an intrinsic ambiguity  
does not exist for potential functionals. Consequently, 
PFAs can be compared point-wise in space to evaluate their accuracy.
To illustrate, we calculate bulk and surface energies 
for a simple case.  

We contrast the direct and variational method of 
calculating the total energy in PFT for a simple model system in Sec.~\ref{s:chisymrel}.
We assess under which conditions the direct evaluation suffices and, 
thereby, examine the symmetry condition on the density-density response function
for existing PFAs. The outcome of this analysis is a PFA to the density-density 
response function for noninteracting, spinless fermions in an arbitrary 
one-dimensional, smooth potential in a box.

In the appendix we show that the direct
and variational evaluation of potential functional TF theory are equivalent. 
We also show what happens as box boundaries, needed in some formal
constructions, are taken far away.  

%%%%%%%%%%%%%%%%%%%%
%%%% BACKGROUND %%%%
%%%%%%%%%%%%%%%%%%%%
\section{Exact statements}
\label{s:exact.pft}

In this section, we compare and contrast PFT with DFT;
this section deals with the exact theory.

\ssec{Basic definitions}

Begin with the variational principle in quantum mechanics
by minimizing over $N$-particle wavefunctions $\Psi$ that are 
antisymmetric, normalized, and have finite kinetic energy:
\ben\label{rrvp}
E[v] = \min_\Psi \left(\brk{\Psi}{\hat T + \hat V\ee + \hat V_v}{\Psi} \right)\,,
\een
where $\hat T$ is the kinetic energy operator,  
$\hat V\ee$ the electron-electron repulsion,
and $\hat V_v$ an external one-body potential,
explicitly denoting the potential as a subscript.

The heart of DFT is the Hohenberg-Kohn (HK) theorem\cite{HK64}
which states, among other things, that 
the ground-state energy of an interacting electronic system can be
found from
\ben
E[v] = \min_\n \left\{ \tilde F[\n] + \int d^3r\, \n(\br)\, v(\br) \right\}\,,
\een
where $\tilde F[\n]$ is a universal functional of the one-electron density $\n(\br)$, 
because it is independent of $v(\br)$.
(We use a tilde to denote density functionals.)
A useful way to define $\tilde F[\n]$ is via the constrained search procedure of
Levy\cite{L79,L82} and Lieb:\cite{Lb82,L83}
\ben\label{FLL}
\tilde F[\n] = \min_{\Psi\to\n} \brk{\Psi}{\hat T + \hat V\ee}{\Psi}\,,
\een
which follows from \Eq{rrvp} by writing the minimization in a two-step procedure,
where the search is performed  over all $\Psi$ yielding the density $n(\br)$.
The original work\cite{HK64} was limited to non-degenerate ground states and
assumed that most reasonable densities would be ground-state densities 
of some interacting electronic problem. 
%\cite{L82,Lb82,L83,EE83,CCR85}  
The constrained search approach is a natural way around these difficulties.

But consider instead\cite{YAW04}
\ben\label{Fv}
F[v] = \brk{\Psi[v]}{\hat T + \hat V\ee}{\Psi[v]}\,,
\een
where $\Psi[v]$ is the ground-state wavefunction of potential $v(\br)$.
Clearly, $\Psi$ is independent of any constant in the potential, and all
potential functionals are functions of the particle number $N$
(for ease of notation, we do not denote this explicitly).
We define it only for those potentials on which we wish to do
quantum mechanics;  a practical choice is the Hilbert space $L^{3/2}+L^\infty$, where
we have a well-defined ground-state wavefunction and energy\cite{DG90}.
If, in addition, we denote the ground-state density as a functional
of the potential,  $\n[v](\br)$, and the dual density functional,
$\tilde v[\n](\br)$, then
\ben\label{Fduals}
\tilde F[\n]=F[\tilde v[\n]],~~~~~F[v]=\tilde F[\n[v]]\ .
\een
In PFT, we can evaluate the ground-state energy directly:
\ben\label{ED}
E[v] = F[v] + \int d^3r\ \n[v](\br)\, v(\br).
\een
Or, we can derive a variational principle in PFT
by minimizing the expectation value of the total energy
over trial potentials $\tv$:
\ben\label{EVP}
E[v] = \min_{\tv} \left(\brk{\Psi[\tv]}{\hat T + \hat V\ee + \hat V_v}{\Psi[\tv]} \right)\ .
\een
With the universal potential functional\cite{YAW04}
defined in \Eq{Fv}, we obtain
\ben
E[v] = \min_{\tv} \left\{ F[\tv] + \int d^3r\ \n[\tv](\br)\, v(\br)\right\}\,,
\label{Ev}
\een
where in the exact case 
the minimizing trial potential is the true external potential $v(\br)$.
Because the right-hand functional is minimized, 
stationarity requires the functional derivative vanish, i.e.,
\ben\label{euler}
\frac{\delta F[v]}{\delta v(\br)}
= -\int d^3r'\ v(\br')\, \chi[v](\br',\br),
\een
where $\chi(\br,\br')=
\delta n[v](\br)/\delta v(\br')$ is the density-density
response function. This is an important exact relation 
between $F[v]$ and $\n[v](\br)$.  Unfortunately, 
the relation is between functional derivatives, not the
functionals themselves.  

However, we can functionally integrate in
several ways.  The simplest is to use 
a coupling constant in the one-body potential:
\ben\label{potcoup}
v\l[v](\br)  = (1-\lambda)\, v_0(\br) + \lambda\, v(\br)\,,
\een
where $0 \le \lambda \le 1$, and 
$v_0(\br)$ is some reference potential. 
Employing the integral form of the Hellmann-Feynman theorem, 
we obtain
\ben\label{EHF}
E[v] = E_0 + \int_0^1 d\lambda\, \int d^3r\
      \n[v\l[v]](\br)\, \Delta v(\br)\,,
\een
where $\Delta v(\br) = v(\br) - v_0(\br)$ is the 
difference of the true and the reference potential.
Choosing a constant reference potential, here $v_0(\br)=0$, 
the universal functional becomes
\ben\label{Fcc}
\mathcal{F}\cc_{\n}[v] = \int d^3r\; \left\{ \bar{\n}(\br)-\n[v](\br) \right\}\, v(\br) \,,
\een
where $\bar{\n}(\br) = \int_0^1 d\lambda\ \n[v^\lambda](\br)$ denotes
the average of the density over the coupling-constant. 
We call $\mathcal{F}\cc$ a \emph{ffunctional} of $\n[w](\br)$, 
because it maps a functional (here $\n[w](\br)$, where $w$ denotes a function of $\br$) 
to another functional, the universal potential functional  
(see appendix~\ref{a:ffunctional} for further discussion).
The gist of \Eq{Fcc} is that knowledge of the potential functional $n[v](\br)$ 
uniquely determines the universal functional $\mathcal{F}\cc$.\cite{CLEB11}

The exercise of checking that $\mathcal{F}\cc$ as constructed by \Eq{Fcc}
satisfies \Eq{euler} will prove useful. 
We take the functional derivative of \Eq{Fcc} 
with respect to the potential $v(\br)$.
This yields 
\bea
\frac{\delta \mathcal{F}\cc}{\delta v(\br)}
&=& -n[v](\br) -\int d^3r'\ v(\br')\,\chi[v](\br',\br)\\
&+& \int_0^1 d\lambda \int d^3r'\ \chi[v\l[v]](\br',\br)\;
\frac{dv\l[v](\br')}{d\lambda}\nonumber\\
&+& \int_0^1 d\lambda \int d^3r'\ n[v\l[v]](\br') \frac{d}{d\lambda}
\frac{\delta v\l[v](\br')}{\delta v(\br)}\nonumber\,,
\eea
which satisfies
Eq.~(\ref{euler}) if, and only if,
\bea
\n[v](\br&)=&\int_0^1 d\lambda\int d^3r'\ \Bigg\{ \chi[v\l[v]](\br',\br)\;
    \frac{d v\l[v](\br')}{d\lambda}\nonumber\\
&+& \n[v\l[v]](\br')\,
    \frac{d}{d\lambda}\frac{\delta v\l[v](\br')}{\delta v(\br)} \Bigg\}.
\eea
This condition is true in turn\cite{CLEB11}, if and only if,
the density-density response function
is symmetric under exchange of coordinates:
\ben\label{symrel}
\chi[v](\br,\br') = \chi[v](\br',\br)\,,
\een
which is an important condition on $\n[v](\br)$ and
is satisfied by the exact density-density response function, by virtue
of it being a second derivative of the ground-state energy.

\ssec{The dual of the Lieb functional}

A more general form of the universal density functional
was constructed by Lieb,\cite{L83} using the Legendre transform of the energy.
Its domain was extended to include any non-negative densities that integrate to 
a given particle number, and is a bifunctional of a potential and a density.
This bifunctional is {\em not} used in PFT and in fact, the
entire procedure is unnecessary in PFT, because a detour 
via a density-potential mapping is unneeded.  However, we show here
its dual in PFT, to illustrate the differences between potential
and density functionals, and to make clear the distinction
with the Lieb construction.

Lieb begins by defining a bifunctional of any pair $n$ and $v$:
\ben\label{Ldef}
L[n,v]=E[v] - \int d^3r\ n(\br)\,v(\br),
\een
Lieb's density functional is then defined as:\cite{L83,NP82}
\ben\label{FLsup}
\tilde F_{\sss L}[n] = \sup_v L[n,v]\,.
\een
How is this related to the potential functional definition
of the universal functional?
Define the PFT dual of $L[n,v]$:
\ben\label{Lnv}
\tilde L[n_1,v_2] = \tilde E[n_1] - \int d^3r\ n[v_2](\br)\, \tilde v[n_1](\br)\,,
\een
where $\tilde v[\n](\br)$ denotes the ground-state potential as a functional of $\n(\br)$.
Starting from the variational principle, we find
\ben
\brk{\Psi[v_2]}{\hat T + \hat V\ee + \hat V_{v_1}}{\Psi[v_2]} \ge E[n_1]
\een
leading to
\ben\label{Fvsup}
F[v] = \sup_n \tilde L[n,v]\,,
\een
because in the exact theory $F[v]=\mathcal{F}\cc_{n}[v]$.
This is the complement to Lieb's definition 
of the universal density functional in the context of PFT,
but the much simpler definition of \Eq{Fv} suffices.
\begin{figure}[htb]
\begin{center}
\includegraphics[angle=0,width=8cm]{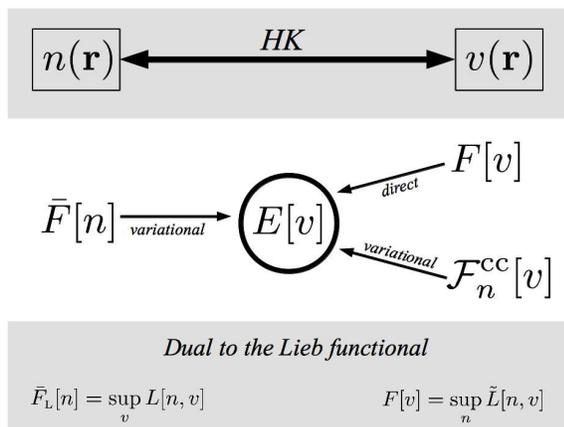}
\end{center}
\caption{\co
Illustrating the relation between the universal functionals
in DFT (left) and PFT (right).}
\label{f:diagram}
\end{figure}

To summarize this introduction to the exact theory,
we illustrate the relation between DFT and PFT in \Fig{f:diagram}.
On the left we depict how the ground-state energy for the
potential $v$ is determined in DFT via $\tilde F[n]$ 
by minimizing over trial densities.
In PFT, on the other hand, the ground-state energy is found either directly
with the given functionals $F[v]$ \emph{and} $n[v]$ via \Eq{ED},
or variationally by sole knowledge of $n[v]$ via $\mathcal{F}\cc_{n}[v]$.

\ssec{Noninteracting systems}

Consider a system of fermions 
in some external potential $v(\br)$  which do {\em not}
interact with one another.
We use a subscript $s$ to denote quantities and functionals for such a system,
and $F[v]$ reduces to $T\s[v]$.   The variational principle simplifies to:
\ben
E[v] = \min_{\tv} \left\{ T\s[\tv] + \int d^3r\ \n\s[\tv](\br)\, v(\br)\right\}\,,
\label{Evs}
\een
and the coupling-constant expression is:
\ben\label{Tscc}
\mathcal{T}_{{\sss S},\n_{\sss S}}\cc[v]
= \int d^3r \left\{ \bar{n}\s(\br) - \n\s[v](\br) \right\}\, v(\br)\ .
\een
The consquence of \Eq{Tscc} is 
that only the knowledge of the noninteracting 
density, $n\s[v]$, is required to determine 
the noninteracting kinetic energy $T\s$.

An alternative expression is given in terms of the virial theorem
for the noninteracting kinetic energy:\cite{SLBB03} 
\ben
\nabla^2 t\s(\br) = -\frac{d}{2}\;\nabla\{ n(\br)\,\nabla v\s(\br) \}\,,
\een
where $d$ is the dimensionality of space
and $t\s(\br)$ the kinetic energy density,
such that $T\s=\int d^3r\ t\s(\br)$.

\ssec{Kohn-Sham scheme}

Up to here we discussed the potential functional analog 
of HK-type density functional theory, where the 
knowledge of the density suffices to determine the total 
energy of either an interacting or noninteracting
system by evaluation of the corresponding potential functionals, 
such as the universal functional and 
the functional for the noninteracting 
kinetic energy on the density. 

\begin{figure}[htb]
\begin{center}
\includegraphics[angle=0,width=7cm]{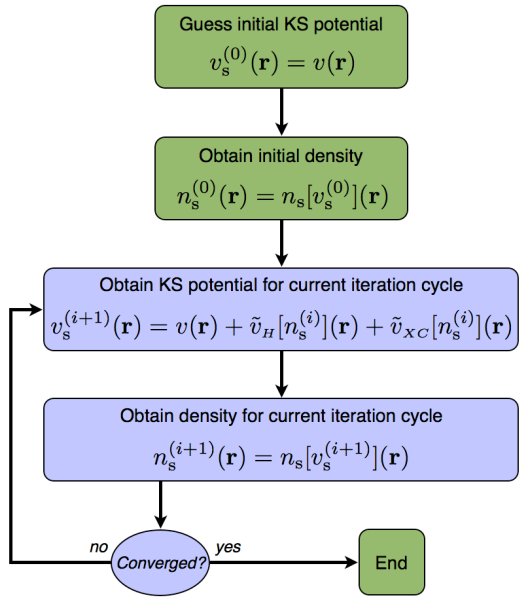}
\end{center}
\caption{\co  
Self-consistent cycle in a PFT calculation within the
KS scheme: In contrast to a KS calculation in DFT, here in PFT,
there is no need to solve the KS equations. 
The iteration begins by guessing a KS potential and
obtaining the initial density via  
$\n\s^{(0)}(\br) = \n\s[v\s^{(0)}](\br)$.
Evaluating \Eq{vspft} on the initial density,
$\n\s^{(0)}(\br)$, 
yields the KS potential of the next iteration, $v\s^{(1)}(\br)$.
The corresponding density is obtained by  
$\n\s^{(1)}(\br) = \n\s[v\s^{(1)}](\br)$,
which is needed to compute the KS potential of the next iteration.
This process is continued until convergence is achieved.}
\label{f:PFTscf}
\end{figure}

But also the noninteracting case can be utilized to yield
the total energy of an interacting system of electrons. 
This is achieved via the celebrated KS scheme.  In what follows 
we describe how PFT could be employed in  the KS construct.
The interacting system is mapped onto a noninteracting
system, requiring that both have the same density. This mapping
is achieved by the KS potential, 
\ben\label{vspft}
v\s(\br) = v(\br) + \tilde v\H[\n\s[v\s]](\br)
+ \tilde v\xc[\n\s[v\s]](\br)\,,
\een
mimicking all many-body interactions
among the electrons in the usual KS-DFT sense 
via the Hartree and XC potentials:
\bea 
\tilde v\H[\n](\br) &=& \frac{\delta \tilde U[n]}{\delta \n(\br)}
= \int d^3r'\ \frac{\n(\br')}{|\br-\br'|}\,,\\
\tilde v\xc[\n](\br) &=& \delta \tilde E\xc[n]/\delta \n(\br)\ .
\eea
With a potential functional to the noninteracting density
$\n\s[v\s](\br)$, which is identical to the interacting density
$\n(\br)$ for the exact KS potential $v\s(\br)$, 
\Eq{vspft} can be solved by standard iteration techniques, 
{\em bypassing} the need to solve the KS equations. 
The process of a KS-PFT calculation is illustrated in \Fig{f:PFTscf}.
A given $\n\s[v\s]$ removes the need
for solving any differential equation in each iteration.

At the end of this iterative process we determine 
the total energy of the interacting electronic system via
\bea\label{EKS}
E[v] &=& \mathcal{T}_{{\sss S}, \n\s}\cc[v\s] + \tilde U[n\s[v\s]]\nonumber \\
     &+& \tilde E\xc[n\s[v\s]] + \int d^3r\ n\s[v\s]\,v(\br)\ .
\eea

%Due to the duality between density and potential functionals
%the existence of the Hartree and XC energy in terms of
%potential functionals is guaranteed, i.e.,
%\bea
%U[v] = \tilde U[n[v]]\,,&~~~& \tilde U[n\s] = U[\tilde v\s[n\s]]\,,\\
%E\xc[v] = \tilde E\xc[n\s[v\s]]\,,&~~~& \tilde E\xc[n\s[v\s]] = E\xc[\tilde v\s[n\s]]\,,
%\eea
%where $\tilde U[n]~\!\!=\!\!\!\!~\int d^3r\; n\s[v](\br)\,v\H[n\s[v\s]](\br)/2$. 
Both the Hartree and the XC contribution can be evaluated readily 
for a given $n\s[v\s](\br)$.  The only missing ingredient for evaluating
\Eq{EKS} is the noninteracting kinetic energy.  But we can 
use our result from the previous section. The noninteracting 
kinetic energy of the KS electrons is given via \Eq{Tscc},
where in that expression the external potential $v(\br)$ becomes 
the KS potential $v\s(\br)$, i.e.,
\ben\label{Tsccvs}
\mathcal{T}_{{\sss S},\n\s}\cc[v]
=\!\! \int\!\! d^3r \left\{ \bar{n}\s(\br) - \n\s[v\s](\br) \right\} v\s(\br)\ .
\een
The knowledge of $\n\s[v\s](\br)$, 
which produces the corresponding $v\s(\br)$ self-consistently, 
suffices to determine the noninteracting kinetic energy of the KS system.

\section{Approximations}
\label{s:app.pft}

In practice, the HK-type of DFT 
requires an approximation 
to the universal functional, $\tilde F\A[n]$, 
where the superscript $\rm A$ denotes an approximation,
such as that used in TF theory, as discussed in
the next section.
Via the variational principle we then 
obtain a relation that determines the density for a
given $v(\br)$:
\ben
\frac{\delta}{\delta n(\br)} 
\left( \tilde F\A[n] + \int d^3r\ \left\{v(\br) - \mu\right\} n(\br) \right) = 0\,,
\een
where the Lagrange multiplier $\mu$ is identical to 
the chemical potential.
This yields an integro-differential equation in
$\n(\br)$ which is typically solved self-consistently,
producing $\n\A(\br)$, whose details
depend on the choice of approximation $\tilde F\A[n]$.

On the other hand, PFT works in a very different way.
In the most general case, 
we have
an approximation to the pair 
$\{n[v](\br), F[v]\}$, 
to obtain the total energy of the many-body 
quantum system directly via \Eq{ED};
in the direct evaluation we need a PFA
to both the density \emph{and} the universal functional as a functional 
of the external one-body potential.  This was the approach used 
in Refs.~\onlinecite{ELCB08,CLEB10},
but takes no advantage of the exact conditions derived 
in Ref.~\onlinecite{CLEB11}.
The semiclassical approach developed in those works\cite{ELCB08,CLEB10}
yields more accurate densities
than kinetic energies, due to the need to take two spatial derivatives
to calculate a kinetic energy and furthermore, is much easier for densities
than kinetic energies, because expansions need only be performed to 
a lower order.\cite{CLEB10}

To take advantage of the results of the previous section, we now
discuss their logic when applied to approximate calculations.
If a pair of approximations
$\left\{F\A, \n\A\right\}$ satisfy \Eq{euler} at $v(\br)$,
then Eqs.~(\ref{ED}) and (\ref{Ev}) yield identical results.
Then no minimization procedure is needed.
But this is {\em not} guaranteed
{\em a priori} in approximate PFT.   Thus, there seem to be
two obvious disadvantages of PFT. 
First, we need to approximate
the density \emph{and} the universal functional separately.
Second, to take advantage of \Eq{Ev}, 
we need to know whether a given pair 
$\{n\A[v](\br), F\A[v]\}$ satisfies the variational principle. 

The functional integration 
in terms of the coupling-constant\cite{CLEB11} removes one of these problems.
With a PFA to the density, $n\A[v](\br)$,
the conjugate approximation for the universal functional follows
from \Eq{Fcc} and reads
\ben\label{FAcc}
\mathcal{F}\cc_{\n\A}[v] = \int d^3r \left\{ \bar{n}\A(\br)-\n\A[v](\br) \right\}\; v(\br) \ .
\een
The important point to note is that only one approximation,
namely $n\A[v](\br)$ is required to 
uniquely determine an approximation to the universal functional.
As a result, we obtain the reverse of the common procedure in DFT: 
In variational PFT we \emph{first} specify which PFA we use
for $n\A[v](\br)$, which \emph{then} determines 
the corresponding $F\A[v]$ via \Eq{FAcc}. 

But this does not automatically cure the second problem.   An approximate
pair constructed in this way does {\em not} automatically satisfy the variational
principle.  It is not even clear which method of calculation (direct, or using the
variational principle) would yield a more accurate answer for a given 
PFA to the density.  However, the previous
section shows that a sufficient condition is the
symmetry condition in \Eq{symrel}, that guarantees identical results, eliminating
the need to perform the minimization.

The utility of this approach 
for calculations on interacting systems is probably limited in practice,
for the same reason that TF theory is largely abandoned in favor of the KS
scheme.  Pure PFT requires
a sufficiently accurate approximation to 
the density of \emph{interacting} electrons 
as a functional of the external one-body potential, to produce approximate energetics
that are accurate enough to bind molecules and generally compete for accuracy
with KS calculations with present XC approximations.
 
A much more likely application of these results is
for noninteracting electrons in a KS potential.
All previous, general statements for the interacting
case analogously apply to the noninteracting case.
First recall the standard procedure in KS-DFT: 
The XC energy is approximated, and
the KS equations are solved self-consistently.
In KS-PFT, however, we additionally need a PFA 
to the noninteracting density, 
which is a less complicated object to approximate 
than its interacting counterpart. 
Then the self-consistent cycle, shown in \Fig{f:PFTscf},
is solved; any exisiting approximation to the
XC energy can be employed in \Eq{vspft}.
The total energy of the many-body quantum
system is finally extracted from \Eq{EKS},
where the noninteracting kinetic energy of KS electrons
is calculated via the conjugate approximation to \Eq{Tscc}:
\ben\label{TsAccvs}
\mathcal{T}_{{\sss S}, \n\s\A}\cc[v]
= \int d^3r \left\{ \bar{n}\s\A(\br) - \n\s\A[v](\br) \right\}\, v\s[v](\br)\ .
\een
Note that the only 
approximation needed to perform this self-consistent
KS-PFT calculation, is $\n\s\A[v](\br)$ (besides the
approximation to $E\xc$, which is also required in
KS-DFT); this scheme is by several orders of magnitude
more efficient than a standard KS-DFT calculation, 
because the KS equations never have to be solved.
However, the applicability of KS-PFT crucially depends on the
accuracy of $\n\s\A[v](\br)$; a major fraction of
the total energy is kinetic, such that only 
tiny errors are allowed. Nevertheless, highly 
accurate approximations to $\n\s\A[v](\br)$ that
satisfy this restriction have already
been derived for model systems\cite{ELCB08,CLEB10}
and approximations for more realistic external potentials
are in development.\cite{LCEB12,CPPG12}

\section{Thomas-Fermi theory: an illustration}
\label{s:TF}

In this section we show how 
PFAs work. 
We use TF theory for this illustration,
because its simplicity in treating the electron-electron
interaction makes the presentation explicit. 
First we recall the usual density functional formulation of 
TF theory for interacting electrons.
Then, we formulate its potential functional counterpart
and confirm that both approaches yield the same result.

%confirm that the approximate pair in TF theory $\{n\TF[v\s], F\TF[v\s]\}$ 
%is variational, and show that the coupling-constant method, 
%which determines  $F\cc[v\s]$ via $n\TF[v\s]$, 
%consistently imposes the variational principle.  
%In the course of showing the latter, 
%we demonstrate and take advantage of the fact 
%that the Hellmann-Feynman theorem holds 
%for variational potential functional theories.

\ssec{Density functional approximation}
\label{sss:TFdftpft}
In the density functional formulation, we need simply an approximation
to $F[\n]$, which in TF theory, is
\ben
\tilde F\TF[\n] = \tilde T\s\TF[\n] + \tilde U[\n]\,,
\een
which is the sum of the local approximation
for the kinetic energy of a noninteracting uniform gas,
\ben
\tilde T\s\TF[\n] = 
\frac{3}{10}\,(3\pi^2)^{2/3}
\int d^3r\ \n(\br)^{5/3}\,,
\label{TTF[n]}
\een
and 
the Hartree energy,
\ben
\tilde U[\n] = \half \int d^3r\ \n(\br)\,\tilde v\H[\n](\br)\,,
\een
where
\ben
\tilde v\H[\n](\br) = \frac{\delta U}{\delta\n(\br)}=
 \int d^3r'\ \frac{\n(\br')}{|\br-\br'|}\ .  
\een
This yields the TF energy functional
\ben\label{ETF[n]}
E\TF[v] = \min_{\n}\left\{\tilde F\TF[\n] +
\int d^3r\ \n(\br)\, v(\br)\right\}\ . 
\een
To find the minimizing density, we functionally differentiate, yielding
an Euler equation for the self-consistent TF density:
\ben
\n\TF(\br) =
\frac{1}{3\pi^2}\left\{2[\mu-v(\br)-\tilde v\H[n\TF](\br)]\right\}^{3/2}\ .
\een 
The density is taken to vanish whenever the argument on the right is
negative, and the chemical potential chosen via normalization:
\ben
\int d^3r\ \n\TF(\br) = N\ .
\een

Lastly, we note that this can always be interpreted in terms
of the KS scheme.  The TF theory ignores XC contributions, so that
\ben
\tilde v\TF\s[n](\br) = v(\br) + \tilde v\H[n](\br),
\een
and the density satisfies:
\ben\label{nTFvs}
\n\TF(\br) = \frac{1}{3\pi}\{2[\mu-\tilde v\TF\s[\n\TF](\br)]\}^{3/2}\ .
\een
%\ben\label{poissoneqn}
%-4\pi\, \n\s\TF(\br) = \nabla^2 v\H[\n\s\TF](\br)\,,
%\een

\ssec{Potential functional approximations}

We now show how TF theory can be derived as a PFA.
Although the final equations are identical, their
derivation is very different.

Because interaction effects are less explicit in PFT, we begin
with an analysis of the noninteracting case.   All PFAs
start with the density as a functional of the potential.  
For plane waves of an extended system with constant potential $v$, one finds
\ben\label{nvbox}
\n\s(v)
= \frac{1}{3\pi^2}[2(\mu-v)]^{3/2}\ .
\een
This then leads to the TF approximation in PFT for noninteracting
electrons in $v(\br)$:
\ben\label{nTFvpft}
\n\s\TF[v](\br) 
= \frac{1}{3\pi^2}\{2[\mu-v(\br)]\}^{3/2}\ .
\een
The same plane waves yield a kinetic energy density in the box:
\ben\label{tvbox}
t\s(v) = \frac{1}{10\pi^2}\,[2(\mu-v)]^{5/2}\,,
\een
which produces the TF PFT for $T\s$:
\ben\label{TTFvpft}
T\s\TF[v] = 
\frac{1}{10\pi^2}
\int d^3r\ \{2[\mu-v(\br)]\}^{5/2}\ .
\een
In fact, insertion of \Eq{nTFvpft} into \Eq{TTFvpft}, eliminating $v(\br)$, produces the
usual TF DFT.\cite{DG90}  Thus the duality shows that knowledge
of the density functionals produces the corresponding potential
functionals, and vice versa:
\ben
T\s\TF[v] = \tilde T\s\TF[\n\s[v]]\,,~~~~~ 
\tilde T\s\TF[\n\s] = T\s\TF[\tilde v[\n\s]]\ .
\een

Armed with this pair of approximations, we can either perform a
direct evaluation, 
\ben\label{ETFdir}
E\TF\dir[v] = T\s\TF[v] + \int d^3r\ \n\s\TF[v](\br)\,v(\br)\
\een
or a minimization
\ben\label{ETFvar}
E\TF\var[v] = \min_{\tv} \left\{ T\s\TF[\tv] + \int d^3r\ \n\s\TF[\tv](\br)\,v(\br) \right\};
\een
and we do not know a priori 
if these yield the same result. In appendix~\ref{a:TFconfirmVP}
we show that for TF theory, in fact, 
\Eq{ETFdir} and (\ref{ETFvar}) are equivalent.

Alternatively, we can use our density PFA 
to {\em construct} a kinetic energy functional 
via \Eq{Tscc}.  Applying this leads to:
\ben\label{TsccTFpft}
\mathcal{T}_{{\sss S}, \n\s\TF}\cc[v]
= \int d^3r \left\{ \bar{\n}\s\TF(\br) - \n\s\TF[v](\br) \right\}\, v(\br)\,,
\een
which is identical to $T\s\TF[v]$. This can be shown in the following way: 
Begin with \Eq{ETFdir} and show that it yields the same energy, when
\Eq{TsccTFpft} is used, i.e., we want to show the equality
\ben\label{EccETF}
E\TF\dir[v] = \mathcal{T}_{{\sss S}, \n\s\TF}\cc[v] + \int d^3r\; \n\s\TF[v](\br)\,v(\br)\ .
\een
Introducing a coupling constant as in \Eq{potcoup} 
(with $v_0(\br)=0$), we can write
\ben\label{ETFac}
E\TF\dir[v] 
= \int_0^1 d\lambda\ \frac{d E\TF\dir[v\l[v]]}{d\lambda}\ .
\een
We further take advantage of the fact that 
the Hellmann-Feynman theorem is satisfied in TF theory\cite{G70}
yielding
\ben
\frac{d E\TF\dir[v\l[v]]}{d\lambda} = \int d^3r\ \n\s\TF[v\l[v]](\br)\, v(\br)\ .
\een
Inserting this into \Eq{ETFac} 
yields the equality in \Eq{EccETF}
via the definition in \Eq{TsccTFpft} and proves the equivalence 
of $\mathcal{T}_{{\sss S}, \n\s\TF}\cc[v]$ and $T\s\TF[v]$.   Thus, from our general
proof, we know that iff our density PFA satisfies the symmetry
condition in \Eq{symrel}, then direct evaluation yields the same as minimization,
so that we can dispense with minimization.
In Sec.~\ref{s:chisymrel}, 
we prove just that in one-dimension, but the proof
is trivial to generalize to three.

From a different perspective, if we did not know $T\TF\s[v]$, our
coupling constant procedure {\em generates} the correct formula.

But full TF theory is about interacting particles, and uses the Hartree
approximation to treat the interaction.  To generate this in PFT,
write the interacting density in terms of the KS potential:
\ben\label{nTFvspft}
\n\TF[v](\br) 
= \frac{1}{3\pi}\{2[\mu-v\s[v](\br)]\}^{3/2}\,,
\een
and
write the Poisson equation in reverse:
\ben\label{poissoneqnpft}
\n\TF[v](\br) = -\frac{1}{4\pi}\,\nabla^2 v\H[v](\br)\,,
\een
where 
\ben\label{vsTFpft}
v\s[v](\br) = v(\br) + v\H[v](\br)\ .
\een
Together, \Eq{nTFvspft} and (\ref{poissoneqnpft}) 
produce the implicit $v$-dependence of $v\H(\br)$,
and so define the TF PFA for the density of interacting electrons.
In fact, the equation for the potential is how TF equations are usually
solved for atoms.\cite{E77,KK81,KTNU55}
In practice, we solve \Eq{vsTFpft}
as illustrated in \Fig{f:PFTscf}. 
The iteration starts by setting $v\H(\br)=0$ in \Eq{vsTFpft},
evaluating $\n\s\TF[v](\br)$ via \Eq{nTFvspft}, and finding
the resulting $v\H[v](\br)$ in \Eq{poissoneqnpft}, 
which is then used to determine the KS potential of the 
next iteration cycle via \Eq{vsTFpft} yielding the
corresponding density via \Eq{nTFvspft}, which inserted into
the left hand side of \Eq{poissoneqnpft} yields
the Hartree potential for the following iteration.  
This process is continued unto convergence.
Finally, the converged KS potential and density are found.  

Then we apply our coupling-constant trick, 
and use as kinetic energy functional 
the TF analog of \Eq{TsAccvs}, i.e., 
\ben\label{TsccTFpftvs}
T_{{\sss S}, \n\s\TF}\cc[v]
= \int d^3r \left\{ \bar{\n}\s\TF(\br) - \n\s\TF[v](\br) \right\}\, v\s[v](\br)\ .
\een
This is almost identical to \Eq{TsccTFpft}, with the only difference
that $v(\br)$ is replaced by $v\s[v](\br)$, which 
is determined self-consistently as just decribed.
This leads to the following PFA to the universal functional,
\ben
\mathcal{F}_{\n\s\TF}\cc[v] = \mathcal{T}_{{\sss S}, \n\s\TF}\cc[v] + U[v]\,,
\een
which is equivalent to $F\TF[v]$.
Again, we can check 
duality:
\ben
U[v] = \tilde U[\n\TF[v]],~~~~
\tilde U[\n] = U[\tilde v\TF[\n]]\ .
\een
Given its symmetry, it does not disturb the symmetry condition, so that
direct evaluation remains sufficient, even in the interacting case.

\section{Asymptotic analysis}
\label{s:asymp}

In this section we apply the theory of the previous sections
to recent suggestions for PFAs.
We restrict the following discussion to noninteracting, 
spinless fermions in a one-dimensional, smooth potential with box boundaries, 
because for this class of potentials an accurate PFA to the density, 
has already been derived\cite{CLEB10,ELCB08} and is of convenient analytical form. 
The derivation produced
$n\s\sc[v](x) = \n\s\sm[v](x) + \n\s\osc[v](x)$, 
where the first term is a smooth, TF-like piece 
and the second an oscillating, quantum correction, 
which are defined as
\bea\label{nsc1dbox}
\n\s\sm[v](x)  &=& \frac{k}{\pi}\,,\\
\n\s\osc[v](x) &=& -\frac{\sin 2\theta}{2\tau\L\,k\sin{\alpha}}\,,
\eea
where we drop the dependency on $x$ 
to preserve a concise notation.  
The quantities in \Eq{nsc1dbox} are
the Fermi wave vector
\ben
k = \sqrt{2(\e\F-v(x))}, 
\een
the classical phase 
\ben
\theta = \int_0^x dx'\, p, 
\een
where $p=\sqrt{2(\e\F-v(x'))}$,
and the classical time for
a particle with energy $\e\F$ 
to travel from $0$ to $x$ or $L$
\ben
\tau = \int_0^x dx'/p,~~~~~
\tau\L = \int_0^L dx/k , 
\een
and the abbreviation
$\alpha = \pi\tau/\tau\L$. 
Note that all quantities in \Eq{nsc1dbox}
are evaluated at the Fermi energy $\e\F$.

In Ref.~\onlinecite{CLEB11} we demonstrated that, 
for generic external potentials which are sufficiently smooth
(such that the basic assumption of the WKB approximation is valid), 
our coupling-constant method combined with
the semiclassical density PFA (derived in 
Ref.~\onlinecite{ELCB08}) yields highly accurate total energies,
almost indistinguishable from the exact answer already
for any $N\ge 2$.  In the following we will analyze the accuracy
of those PFAs, both in terms of energy
and in real space. Then we explain the source
of the observed accuracy via an asymptotic analysis 
in the large-$N$ and the classical continuum limit.\cite{CLEB10}
Furthermore, we calculate the contributions 
to the asymptotic correctness coming from
distinct regions -- the interior and the edge; those
are analogs of the surface and bulk of a real solid.
In the following we present only the essence of our analysis.
We refer the interested reader to the supplemental material\cite{supp}
for further numerical details.

\ssec{Large-$N$ asymptotic expansion of energies}
We analyze the asymptotic behavior of the total energy in the large-$N$ limit
to assess the accuracy of approximations.
We will examine the behavior of two different PFAs, 
the independent semiclassical approximation (ISA) 
and the density-driven semiclassical approximation (DSA).  

The first consists of two independent semiclassical approximations, 
one for the density, first derived in Ref.~\onlinecite{ELCB08},
and another for the kinetic energy density.
This derivation requires expanding to a higher order in $\hbar$, and yields an asymptotically 
correct expansion in the interior, but fails near an edge.
This failure was patched to the asymptotically correct solution near the
edge in Ref.~\onlinecite{ELCB08}, in order to produce a uniformly asymptotic approximation.
A better patching scheme was developed in Ref.~\onlinecite{CLEB10}.

The second, denoted DSA, was first derived in Ref.~\onlinecite{CLEB11}, 
using the coupling-constant construction studied here.  The advantage is 
that one needs only the semiclassical formula for the density alone, 
which is uniformly asymptotic, and so the kinetic energy density derived 
from it by functional integration via the coupling-constant method should automatically
be uniformly asymptotic.   An obvious question
is its performance relative to the ISA, and especially the behavior of both 
near the edge of the box.

\begin{figure}[htb]
\begin{center}
\includegraphics[angle=0,width=9cm]{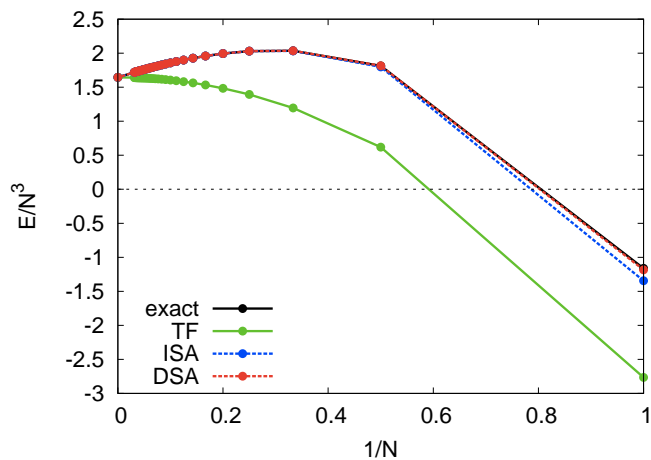}
\end{center}
\caption{\co  
Numerical confirmation of the leading coefficient 
$c_0$ of \Eq{EasymplargeN} from the exact calculation, 
the ISA and DSA for $v(x)=-8\sin^2{(\pi x)}$, 
where the maximum number of particles considered is $N=32$.}
\label{f:E_N}
\end{figure}

In Table~\ref{t:E_N}, we list total energies for several $N$ values 
for a generic external potential $v(x)=-8\sin^2{\pi x}\,,\, 0\le x\le 1$.
The incredibly rapid convergence of the DSA to the exact energy as $N$ grows
is readily apparent. We can understand and quantify this as follows:
\begin{table}[htb]
\caption{Exact total energy and errors in the TF approximation, 
the ISA, and DSA
for $N$ particles in the external potential $v(x)=-8\sin^2{(\pi x)}$, 
where $0\le x\le 1$.}
\label{t:E_N}
\begin{ruledtabular}
\begin{tabular}{ l  c c c c}
\multicolumn{1}{l}{} &
\multicolumn{1}{c}{} & 
\multicolumn{3}{c}{$E\A -E$}\\ \cline{3-5}
$N$ &
$E$ & 
TF  & 
ISA &
DSA \\
\hline
  1 &	    -1.1615 &	    -1.603  &	 -0.1825 &	-0.0221\\
  2 &	    14.510  &	    -9.554  &	 -0.1200 &	 0.0054\\
  4 &	   129.95   &	   -40.78   &	 -0.0357 &	 0.0011\\
  8 &	   972.65   &	  -162.5    &	 -0.0098 &	 0.0002\\
 16 &	  7316.4    &	  -642.8    &	 -0.0026 &	 $2\times 10^{-5}$\\
 24 &	 24082      &	 -1438      &	 -0.0010 &	 $7\times 10^{-6}$
\end{tabular}
\end{ruledtabular}
\end{table}
Expanding the total energy in powers of $N$ yields
\ben\label{EasymplargeN}
E(N)/N^3 = c_0 + c_1/N + c_2/N^2 + c_3/N^3 + c_4/N^4 + \dots\ .
\een
We characterize the accuracy of an approximation\cite{ELCB08}
by measuring the deviation from those exact coefficients. 
We perform this analysis explicitly for the generic external potential, 
assuming the qualitative features are
independent of the specific potential, once it is reasonably smooth.
In \Fig{f:E_N}, we plot $E(N)/N^3$ both exactly and for the various approximations.

Because of the box boundary conditions, the energies approach those
of a flat box as $N$ grows, and in fact, the leading two coefficients
are just $c_0=\pi^2/6$ and $c_1=\pi^2/4$, respectively.  Remarkably, the TF approximation
is only AE0, i.e., it errs in the value of $c_1$, as it does {\em not} recover
the flat box results exactly.\cite{ELCB08}  But the semiclassical
corrections greatly improve over TF theory.  To analyze them, we define
the residual energy, $\Delta E= E - E^{\rm flat}$, where the flat
box result is known analytically.  Then,
\ben
\Delta E(N) = c_2'\, N + c_3 + c_4/N + \ldots\,,
\een
where $c_2' = (c_2 - \pi^2/12)$. 
We find the ISA correctly
reproduces $c_2'=-4$, but makes a small error in $c_3$ (about $0.25\%$). 
Thus this approximation is AE2, as discussed in Ref.~\onlinecite{ELCB08}, 
and is almost AE3.

But the new approximation, the DSA, 
using only the density formula and coupling-constant
integration, is at least AE4.  It is beyond our numerical accuracy to determine
$c_5$ sufficiently accurately, although it appears that even this coefficient
may be exact within the DSA.   Thus, by reproducing two more
terms in the asymptotic expansion exactly, we get tremendous improvements
in accuracy, even at $N=1$.   This illustrates the potential of the coupling-constant
method to generate incredibly accurate approximations.

\ssec{Classical continuum limit of energies}

An alternative limit in which TF also becomes exact was described in Ref.~\onlinecite{CLEB10}.
We define the approach to the classical continuum limit by increasing
the number of particles in a system from its original value $N$ to
$N' > N$, while simultaneously replacing $\hbar$ by $\gamma\hbar$, where
$\gamma=N/N'$.  As $N' \to \infty$,
the energy differences
between discrete eigenvalues becomes infinitesimal 
and a continuum is formed. 
The advantage of this limit, as opposed to large $N$, is that it approaches
the TF solution {\em of the original problem with $N$ particles}, rather
than approaching the $N\to\infty$ problem of the previous section.

Expanding the total energy in powers of $\gamma$ about $0$ yields
the expansion
\ben\label{Egammaexp}
E(\gamma) = E\TF\, (1 + b_1\, \gamma + b_2\, \gamma^2 + \ldots),
\een
where
\ben
b_p = \frac{1}{p!} \frac{d^p E}{d\gamma^p} \Big|_{\gamma=0}\, \Big/ E\TF \ .
\een
Such an expansion is expected to be asymptotic rather than convergent.
We also define $E^{(p)}$ as the sum up to the $p$-th order of such
terms.
We find, for $N=1$, that while $E^{(2)}$ is more accurate than lower-order
truncations, $E^{(3)}$ overshoots.  For $N>1$, all successive terms up to third-order
always improve accuracy.

Under this $\gamma$-scaling, the TF approximation is independent of $\gamma$.
We find both PFAs, the ISA and DSA, reproduce $b_1$ exactly, but neither yields $b_2$.
This is not surprising, as these approximations were derived only to yield
the leading corrections to TF in this limit.  
For $N=1$, we find that neither approximation yields particularly accurate expansion
coefficients and that their absolute errors for the coefficients are comparable.
Regardless, these errors in the coefficients do not translate into an inaccurate energy;
as we showed in the previous section, the DSA energy is very accurate for $N=1$.
As $N$ grows, the errors in the coefficients rapidly shrink, 
but from analyzing the coefficients alone the DSA would appear 
no more accurate than the ISA.
As shown in the previous section, however, the DSA is far more accurate 
and converges more rapidly than the ISA with increasing $N$.  
Therefore the asymptotic expansion in $\gamma$ is not useful for understanding 
the improved accuracy of the DSA relative to the ISA.

\ssec{Analysis in real space}

In density functional approximations, a difficult and vexing issue is the
ambiguity of the energy density of an approximate density functional
for an energy\cite{CLB98,BCL98}.   For example, one can always add the Laplacian of the
density to the energy density of even a local approximation without changing
the functional, since that addition integrates to zero as long as the
density vanishes on the boundary.   This difficulty complicates any point-wise
comparison between approximate functionals\cite{BPW97} and has hampered our
ability to construct improved approximations, especially local hybrids\cite{CLB98,AK12}.

A great advantage of PFT is that such ambiguity does not exist:
PFAs approximate a given choice of exact energy density, and so
can be compared point-wise.  There is no ``gauge'' ambiguity\cite{TSSP08}.
\begin{figure}[htb]
\begin{center}
\includegraphics[angle=0,width=9cm]{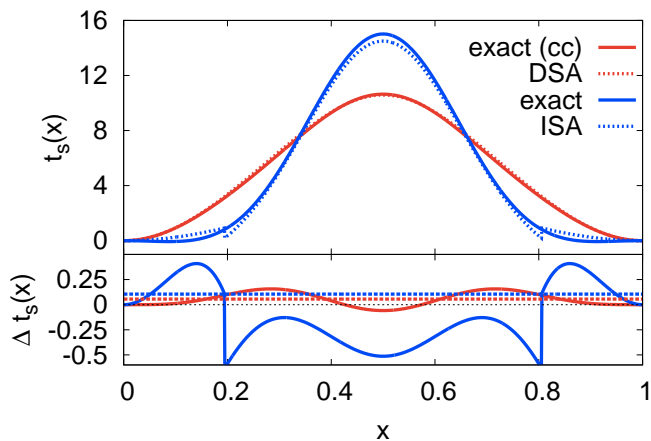}
\end{center}
\caption{\co 
Comparing ISA kinetic energy density (dotted blue)
with its exact counterpart (solid blue), and DSA (dotted red)
with its exact counterpart (solid red) 
for one particle in $v(x)=-8\sin^2{(\pi x)}$.
The lower panel shows the errors, 
indicating average errors by dashed lines.}
\label{f:dts_N1}
\end{figure}
In \Fig{f:dts_N1}, we illustrate this for the single particle in the single well of depth 8.
The ISA approximates one definition of the kinetic energy
density, while the DSA approximates another.  
But in each case, they can be compared for accuracy point-wise 
to the respective exact curve.  Again we see
that the DSA is more accurate everywhere. 
Its maximum errors are much smaller than
those of the ISA, and the average errors are also much smaller.

To check this idea, we have seen above that both semiclassical PFAs
show higher-order asymptotic exactness than the TF approximation.
We now ask if these improvements are visible in separate
spatial regions, not just for quantities integrated over the entire system.
In particular, we know that the limit as $\gamma\to 0$ is different for fixed values
of $x$ (interior) than for fixed values of $\gamma x$, the edge (or surface) region\cite{CLEB10}.
But the semiclassical density approximation 
is supposed to be uniformly asymptotic, i.e., have the
same degree of AE for each region separately.   To test this, we must define a dividing
line between the interior and edge.  We choose
the half-phase point $x_{\pi/2}$,
defined by the following condition 
on the classical phase\cite{ELCB08}:
\begin{equation}\label{halfphasepoints}
\theta(x_{\pi/2}) = \pi/2\ .
\end{equation}
This is our measure to split the box into
an interior ($L/2-|L/2-x|>x_{\pi/2}$) 
and edges (the rest).

This condition has already been used for the boundary-layer
analysis of the ISA in Ref.~\onlinecite{CLEB10}.
As $N$ grows, $x_{\pi/2} \sim 1/N$; or as $\gamma\to 0$, $x_{\pi/2} \sim \gamma$.
We define a surface kinetic energy as the energy in this region, 
and analyze the accuracy of our approximations for this quantity.  

The energy from distinct regions follows the same expansions as \Eq{EasymplargeN} 
and (\ref{Egammaexp}), but with different coefficients $c\inter_p$, 
$c\edge_p$, $b\inter_p$, and $b\edge_p$.
As $N$ grows, the boundary between the edge and interior 
-- the half-phase point $x_{\pi/2}$ --
is shifted towards the edge, such that in the limit $N\to\infty$ the edge region
vanishes and the interior extends over the entire length of the system.  
The same is true for the classical continuum limit.   Hence, the leading order 
coefficient of both expansions close to the edge vanishes, i.e.,
$c\edge_0=0$ and $b\edge_0=0$.  
   Also notice that we need to distinguish between two exact results, when splitting up
the energy into contributions from different spatial regions. This is due to the fact
that we use two different definitions of the kinetic energy density. 
The ISA stems from the Laplacian definition, 
whereas the DSA yields a different definition as illustrated in \Fig{f:dts_N1}.
However, the difference of the energy values between those definitions vanishes
with increasing $N$; therefore we report only exact regional 
energies using the Laplacian definition,
but note that the errors of DSA are calculated with respect to the
exact energy density defined by the coupling-constant method for each region.
\begin{table*}[htb]
\caption{Exact total energy and errors of approximations
(TF, ISA, and DSA) in the interior and close to the edge 
for $N$ particles in the external potential $v(x)=-8\sin^2{(\pi x)}$, 
where $0\le x\le 1$.}
\label{t:delta_E_N}
\begin{ruledtabular}
\begin{tabular}{ l  c c c c c c c c}
\multicolumn{1}{l}{}&
\multicolumn{4}{c}{interior}&
\multicolumn{4}{c}{edge}\\ \cline{2-5}\cline{6-9}
\multicolumn{2}{l}{}&
\multicolumn{3}{c}{$E\A-E$}&
\multicolumn{1}{l}{}&
\multicolumn{3}{c}{$E\A-E$}\\ \cline{3-5}\cline{7-9}
$N$ &
$E$ & 
TF  & 
ISA &
DSA &
$E$ & 
TF  & 
ISA &
DSA\\
\hline
  1	&    -0.6721 &    -1.081 & -0.1051 & 0.0076            &  -0.4891 &  -0.5219 & -0.0774 & -0.0297 \\
  2	&     9.631  &    -8.215 & -0.1505 & 0.0068            &   4.876  &  -1.3361 &  0.0304 & -0.0014 \\
  4	&   108.5    &   -41.28  & -0.0666 & 0.0012            &  21.41   &   0.5091 &  0.0308 & $-6\times 10^{-5}$ \\
  8	&   891.2    &  -178.9   & -0.0199 & 0.0002            &  81.23   &  16.49   &  0.0101 & $-1\times 10^{-6}$ \\
 16	&  7004      &  -738.7   & -0.0054 & $2\times 10^{-5}$ & 310.8    &  96.16   &  0.0028 & $-2\times 10^{-8}$ \\
 24	& 23391      & -1677     & -0.0025 & $7\times 10^{-6}$ & 686.9    &  239.2	 &  0.0013 & $-3\times 10^{-9}$  
\end{tabular}
\end{ruledtabular}
\end{table*}
\begin{figure}[htb]
\begin{center}
\includegraphics[angle=0,width=9cm]{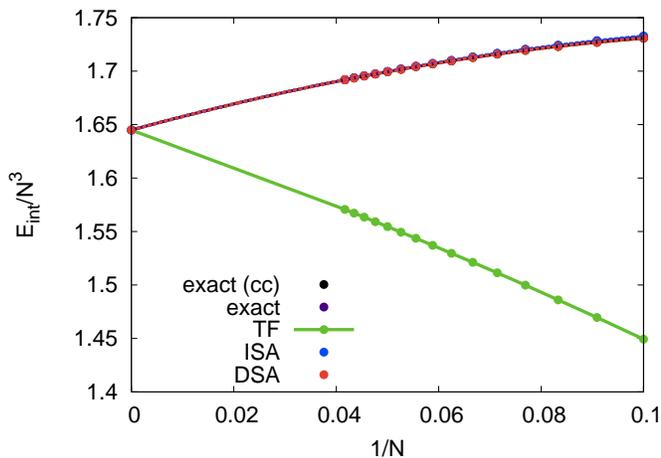}
\end{center}
\caption{\co
Numerical extraction of leading coefficients 
in \Eq{EasymplargeN} in the interior from the exact calculation, 
the ISA and DSA for $v(x)=-8\sin^2{(\pi x)}$, 
where the maximum number of particles considered is $N=24$.}
\label{f:E_N_int}
\end{figure}
First we consider the total energies of the interior given in 
the first columns of Tab.~\ref{t:delta_E_N}. We plot $E\inter(N)/N^3$ 
in \Fig{f:E_N_int}, illustrating the large-$N$ limit, in which 
the energy expansion in the interior approaches $\pi^2/6$. 
This analysis also yields that the ISA correctly reproduces
$c_0$, $c\inter_1 = 1.313$, and $c\inter_2 = -4.492$, but fails to yield 
$c\inter_3$ (with a small error of about $0.1\%$).
On the other hand, the DSA exactly reproduces 
the coefficients up to at least $c\inter_4$.
\begin{figure}[htb]
\begin{center}
\includegraphics[angle=0,width=9cm]{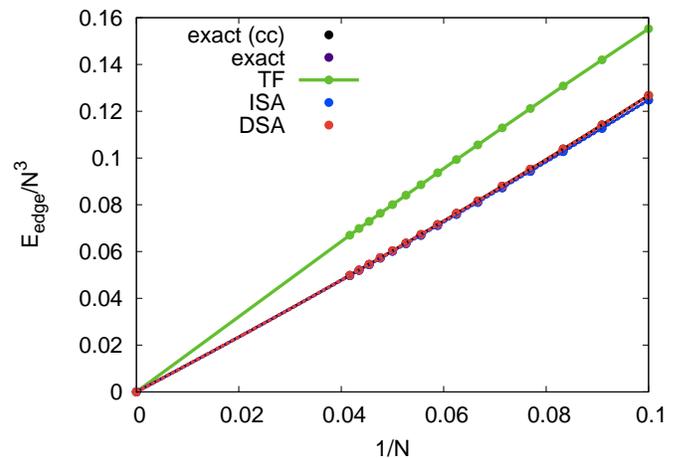}
\end{center}
\caption{\co
Numerical extraction of the leading coefficients 
in \Eq{EasymplargeN} close to the edge from the exact calculation, 
the ISA and DSA for $v(x)=-8\sin^2{(\pi x)}$, 
where the maximum number of particles considered is $N=24$.}
\label{f:E_N_edge}
\end{figure}

Next we consider the edge region, for which the energy contributions
and errors of approximations are listed in each second column 
of Tab.~\ref{t:delta_E_N}. As the size of this region vanishes 
in the limit $N\to\infty$, the leading term in the large-$N$ expansion 
falls off linearly with $N$. This is illustrated in \Fig{f:E_N_edge}. 
In analogy to the interior, the ISA correctly yields $c\edge_1=1.147$ and $c\edge_2=1.16$,
but makes a mistake for $c\edge_3$ by about $0.5\%$. The DSA, however,
yields exact coefficients up to at least $c\edge_4$ and probably also 
for higher order contributions. The numerical accuracy of our analysis is 
not high enough to give a definite answer beyond this order.

In conclusion, we could demonstrate that 
even when we consider energy contributions from separate spatial regions,
the ISA is AE2, whereas the DSA is at least AE3. 
Furthermore this analysis shows that the accuracy achieved by both approximations 
over the entire system is not due to error cancellations in different spatial regions;
it is caused by virtue of both approximations capturing the large-$N$ asymptotic expansion
in separate spatial regions sufficiently accurate.

\section{The importance of being symmetric}
\label{s:chisymrel}
In this section we contrast the direct method of calculating
the total energy in PFT given by \Eq{ED} to the variational 
method in \Eq{EVP}. Determining the total energy variationally
as in \Eq{EVP} is only sensible for calculations that involve
approximations. In the exact case the total energy is always minimized
by the true external potential of the given problem.
We also demonstrate that the particular PFA to the density, $n\s\sc[v](x)$, 
derived in Ref.~\onlinecite{ELCB08}, violates the 
symmetry condition of \Eq{symrel}. 
Consequently, when we perform a search over trials potentials, 
the minimum energy is not given at $\tv(x)=v(x)$ 
as predicted by the variational principle.
\begin{figure*}[htp]
\begin{center}
\subfigure[~$N=1$]{
\includegraphics[width=8.5cm]{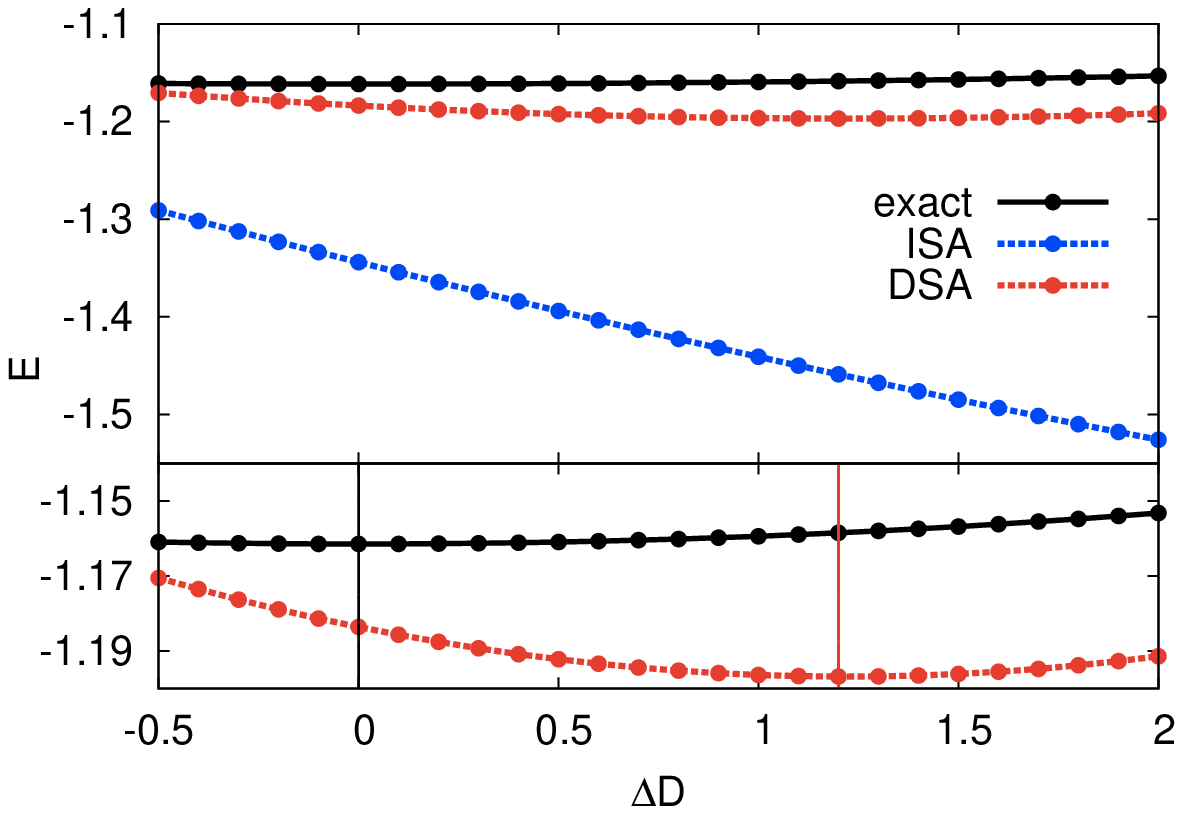}}\quad
\subfigure[~$N=2$]{
\includegraphics[width=8.5cm]{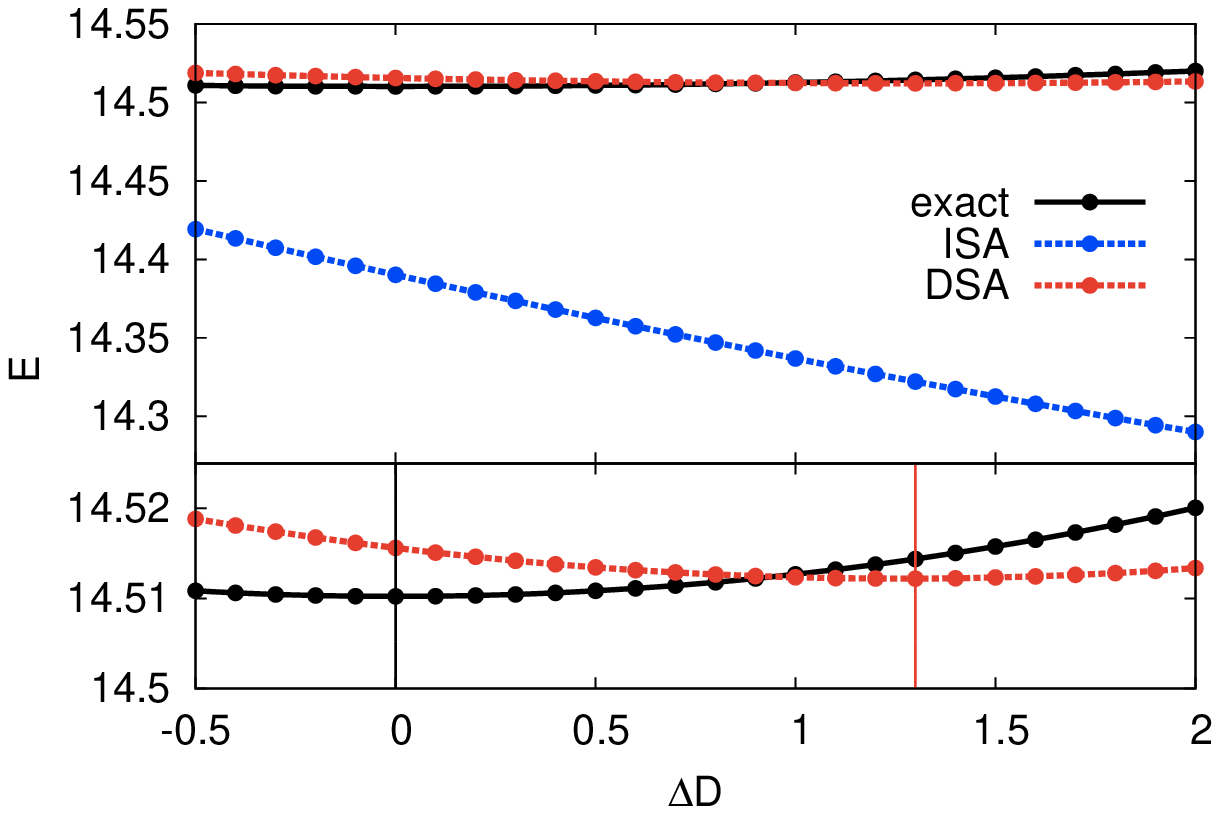}}
\caption{\co
Variational calculation of the total energy (black) in comparison to 
the ISA (blue) and DSA (red) of $N$ noninteracting,
spinless fermions in an external potential   
$v(x)=-8\sin^2(\pi x)$, where the minimization
is performed over trial potentials $\tv(x)=v(x) - \Delta D\sin^2(2\pi x)$. 
The lower panel shows a magnification of the exact result and the DSA,
illustrating the position of the variational minima.}
\label{f:DEsd}
\end{center}
\end{figure*}

To illustrate this we consider $N$ noninteracting, spinless fermions in 
an external, one-body potential
$v(x)=-8\sin^2(\pi x)$ in a box, where $0<x<1$. 
We perform a variational calculation of the total energy, where 
we evaluate \Eq{EVP} on an extremely limited class of trial potentials
$\tv(x)= v(x) - \Delta D\sin^2(2\pi x)$.
The result is illustrated in \Fig{f:DEsd} for $N=1$ and $N=2$.
The black curve depicts the exact calculation with
the minimum located at $\Delta D=0$. The blue curve
corresponds to the ISA, which might not even have a minimum,
because two independent approximations 
$\n\s\A[v]$ and $T\s\A[v]$ 
are employed. 
The red curve is the DSA which, despite being more accurate,
also does not minimize at the true external potential, but 
at $\Delta D=1.2$ for $N=1$ and at $\Delta D=1.3$ for $N=2$. 
\begin{table}[htb]
\caption{Total energy of $N$ noninteracting,
spinless fermions in the external potential $v(x)=-8\sin^2{(\pi x)}$, 
where $0\le x\le 1$ and the absolute errors of the direct evaluation 
within TF, ISA, and DSA. Additionally, we list the absolute errors 
of the variational evaluation of the DSA together with the minimizing trial potential.}
\label{t:Evar}
\begin{ruledtabular}
\begin{tabular}{ l c c c c c c}
\multicolumn{2}{c}{} &
\multicolumn{3}{c}{direct} & 
\multicolumn{2}{c}{variational} \\ \cline{3-5} \cline{6-7}
\multicolumn{2}{c}{} &
\multicolumn{3}{c}{$E\A$} & 
\multicolumn{1}{c}{} & 
\multicolumn{1}{c}{$E\A$} \\ \cline{3-5} \cline{7-7}
$N$ &
$E$ & 
TF  & 
ISA &
DSA &
$\Delta D$ &
DSA \\
\hline
1 &  -1.161 &   -1.603 & -0.183 & -0.022     & 1.2 & -0.038\\ 
2 &  14.510 &   -9.554 & -0.120 &  0.005     & 1.3 & -0.002\\ 
4 & 129.953 &  -40.778 & -0.036 &  0.001     & 0.0 &  0.001\\
8 & 972.652 & -162.496 & -0.010 &  $10^{-4}$ & 0.0 & $10^{-4}$\\
\end{tabular}
\end{ruledtabular}
\end{table}
Furthermore, in Tab.~\ref{t:Evar} we list total energies
given by the exact calculation and absolute errors of several 
approximations (TF, ISA, and DSA) obtained from the direct evaluation
for increasing $N$. Additionally, we also list the absolute errors
of the DSA from the variational calculation along with 
the effective depth $\Delta D$ of the minimizing trial potential.
This analysis shows that 
a minimization over some trial potentials can yield a more accurate 
result at a trial potential different from the true 
external potential, but that this is not always the case; 
for example, the DSA in \Fig{f:DEsd} 
for $N=2$ yields a more accurate energy at its variational minimum 
than at $\Delta D=0$. However, for $N=1$ the energy of the DSA
at the variational is less accurate. Furthermore, as $N$ increases
the error of the DSA decreases quickly, and its variational minimum
coincides with the true external potential as is demonstrated by
the variational results in Tab.~\ref{t:Evar}.  Consequently, for large $N$
a variational minimization of the DSA becomes obsolete, since
the minimum will be given at the external potential of the given problem.
Furthermore, to find the true global minimum (not guaranteed to exist
for a semiclassical expansion, as DSA is), one needs to search over 
{\em all} trial potentials, not just simple multiples of the external
potential, which could be achieved via a set of self-consistent
equations.  

From the previous sections, we know that DSA's failure to
minimize at the correct potential must be because
it violates  
the symmetry condition
\ben
\frac{\delta n\s\sc[v](x)}{\delta v(x')}\Bigg|_N 
= \frac{\delta n\s\sc[v](x')}{\delta v(x)}\Bigg|_N.
\een
The specific PFA 
$n\s\sc[v](x) = n\s\sm[v](x) + n\s\osc[v](x)$,
consists of a smooth ($n\s\sm$) and an oscillating ($n\s\osc$) piece.\cite{ELCB08}
To first order in $\delta v(x)$ the Fermi energy $\e\F$ changes as
\ben
\delta\e\F = \frac{1}{\tau\L}\int_0^L dx\ \frac{\delta v(x)}{k}\,,
\een
the smooth piece of the density yields
\ben\label{chiscsm}
\chi\s\sm(x,x')
= \frac{1}{\pi k}\left[ \frac{1}{\tau\L\,p} - \delta(x-x') \right]\,, 
\een
and the oscillating piece gives
\ben\label{chiscosc}
\chi\s\osc(x,x') = \n\osc(x) \left[ a + b\,\cot{2\theta} - c\, \cot{\alpha} \right]\,,
\een
with
\bea
a &=& \frac{1}{\tau\L\, p} 
\left(\frac{\xi\L}{\tau\L}-\frac{1}{p^2}-\frac{1}{k^2}\right) + \frac{\delta(x-x')}{k^2},\\
b &=& \frac{2}{p} \left[ \beta - \eta(x-x') \right],\\
c &=& \frac{\pi}{\tau\L\, p} 
\left( \frac{\eta(x-x')-\beta}{p^2} + \frac{\beta \xi\L-\xi}{\tau\L} \right),
\eea
where
$\eta(x-x')$ denotes the Heaviside step function,
$\beta = \tau/\tau\L$,
$\xi = -(d\tau/d\e)|_{\e=\e\F}$, and 
$\xi\L = -(d\tau\L/d\e)|_{\e=\e\F}$.
As expected, the functional derivative of the smooth, TF-like piece 
is symmetric under exchange of $x$ and $x'$. However, the functional 
derivative of the oscillating piece, is \emph{not} symmetric. 
This is the reason, why the red curve in \Fig{f:DEsd} 
does not minimize at $v(x)$.   

\begin{figure}[htb]
\begin{center}
\includegraphics[angle=0,width=9cm]{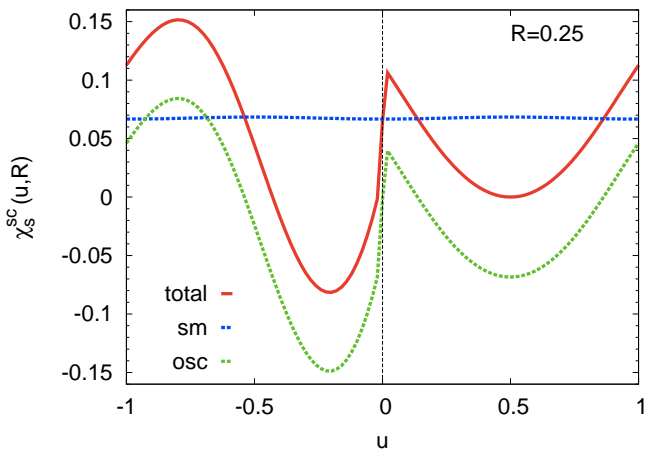}
\end{center}
\caption{\co 
PFA to the static density-density response function for fixed 
average coordinate $R=0.25$ (where $R=(x+x')/2$) 
as a function of the relative coordinate $u=x'-x$
of one particle in the external potential $v(x)=-5\sin^2{\pi x}$, 
where $0\le x\le 1$. This demonstrates explicitly 
that \Eq{chisc} is not symmetric under exchange of coordinates.
In particular, we confirm that the smooth piece given
in \Eq{chiscsm} is symmetric, 
whereas the oscillating piece in \Eq{chiscosc} is not.
}
\label{f:chiscRu_comp}
\end{figure}

\begin{figure}[htbp]
\begin{center}
\includegraphics[angle=0,width=9cm]{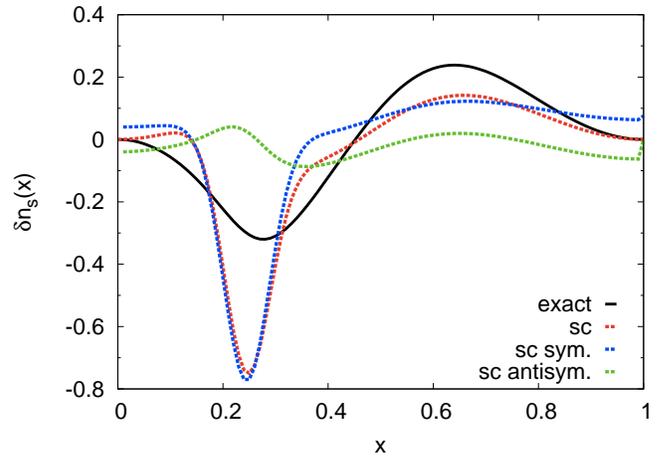}
\end{center}
\caption{\co 
Exact (black) and semiclassical (red) 
change in density of one particle 
in the external potential $v(x)=-5\sin^2{\pi x}$,
due to a change of potential proportional to 
a gaussian of width 0.001 centered at $x=0.25$.
Symmetric (blue) and antisymmetric (green) 
semiclassical contributions are also shown.
}
\label{f:deltan}
\end{figure}

The fall-out of this analysis is an explicit PFA 
to the static density-density response function 
\ben\label{chisc}
\chi\s\sc[v](x,x') 
= \chi\s\sm[v](x,x') + \chi\s\osc[v](x,x')
\een
for noninteracting, spinless fermions in an external potential $v(x)$ 
confined by box boundaries, which is an interesting result in itself.
As an example we consider one particle in $v(x)=-5\sin^2{\pi x}$. 
Introducing average and relatives coordinates, i.e.,  
$R=(x+x')/2$ and $u=x'-x$, we plot \Eq{chisc} 
in \Fig{f:chiscRu_comp}; this explicitly 
demonstrates that \Eq{chisc} is not symmetric under exchange 
of coordinates. Additionally, in \Fig{f:deltan} 
we illustrate the symmetric and antisymmetric contributions 
to the change in density 
$\delta \n\s(x) = \lim_{f\to 0} (n[v+f\,v](x)-n[v](x))/f$ 
calculated via \Eq{chisc} when the external potential 
is perturbed by $\delta v(x) = f\,g(x)$, where 
$g(x) = \exp{[-(x-x_0)^2/(4\sigma)]}/(2\sqrt{\pi\,\sigma})$,
which approaches a Dirac delta function centered at $x_0$
in the limit $\sigma\to 0$.

\section{Conclusion}
In this work, we have established several formal properties of the
potential functionals introduced in Ref.~\onlinecite{CLEB11}, 
especially in terms of their duality to density functionals.\cite{YAW04}
We have also shown that the methodology of Ref.~\onlinecite{CLEB11}
can be employed to produce more accurate potential functionals
than any that had previously existed.  
This higher accuracy can be attributed to the ``unreasonable
utility of asymptotic expansions'',\cite{S80,S81} because the
new coupling-constant procedure considerably improves the accuracy
of the asymptotic expansion in powers of $1/N$, where $N$ is the
particle number.

Of course, the major drawback of this line of investigation remains:
Only in one dimension, and then only for smooth potentials confined by
box boundaries, do we have explicit expressions that are uniformly more accurate than
TF theory, at the present time.  The thrust of the present investigation
is to show how promising such approximations are, and to dangle the
hope of tremendous improvement over present day density functional
approximations, especially for orbital-free calculations.

%%%%%%%%%%%%%%%%%%%%
%% ACKNOWLEDGMENT %%
%%%%%%%%%%%%%%%%%%%%

%We gratefully acknowledge funding from NSF under 
%grant number CHE-0809859.
\section{Acknowledgments}
K.B. and A.C. acknowledge support 
by NSF under Grant No. CHE-1112442.
A.C. also acknowledges Lucas Wagner, 
Stefano Pittalis, and Raphael Ribeiro 
of the Burke group for useful comments 
and proofreading of the manuscript.

\appendix

\section{Confirming the variational principle 
in potential functional Thomas-Fermi theory}
\label{a:TFconfirmVP}

We show that in noninteracting TF theory the direct evaluation 
of the total energy in \Eq{ETFdir} yields the same 
as the minimization over trial potentials given in 
\Eq{ETFvar}.  If the approximate pair
$\{ \n\s\TF[\tv](\br), F\TF[\tv] \}$ satisfies 
the variational principle, i.e., 
\ben\label{a:vptfpft}
\frac{\delta}{\delta \tv(\br)} 
\left( F\TF[\tv] 
+ \int d^3r\ \left\{v(\br) - \mu\right\} \n\s\TF[\tv](\br) \right) = 0\,,
\een
then the statement above is true.
To confirm the latter, 
define the local chemical potential 
$\tmu[v](\br) = \mu - v(\br)$,
which is directly related to the density via
\ben  
\n\s\TF[v](\br) = \frac{\{ 2\tmu[v](\br) \}^{3/2}}{3\pi^2}\ .
\een
Take the functional derivative in \Eq{a:vptfpft} using 
the chain rule, e.g.,
\ben
\frac{\delta T\s\TF}{\delta \tv(\br)}
= \int d^3r'\ \frac{\delta T\s\TF}{\delta \tmu(\br')} 
\frac{\delta \tmu(\br')}{\delta \tv(\br)}\,,
\een
considering that
\ben
\frac{\delta \tmu(\br')}{\delta \tv(\br)}
= -\delta(\br'-\br)\,,
\een
where $\delta(\br'-\br)$ denotes the Dirac delta function,
and keeping the chemical potential $\mu$ fixed, 
since it is determined at the 
end of the minimization by requiring normalization.
Then, we find that
$\tv(\br) = v(\br)$,
as expected.

\section{Walls at infinity}
The DSA can be applied to potentials, 
for which $v(x) \to 0$ as $|x|\to\infty$, and
where $v_0=0$ is chosen as a reference potential. 
However, for this choice the coupling-constant integral in Eqs.~(\ref{Fcc}) 
and (\ref{Tscc}) might be undefined, because the particle number 
may change abruptly with the coupling constant. To assure 
the existence of the coupling-constant integral,
we use the formal device of introducing hard walls and taking
the limit of infinite separation at the end of the calculation.

We demonstrate this procedure for a simple case.
Consider $v(x)$ to be a finite square well.
Assume the depth of the square well is such 
that there are two bound states. When the coupling-constant
integral is performed the depth of the well changes from its initial
value at $\lambda=1$ to zero at $\lambda=0$. As the depth decreases,
there is a point where the energy of the second bound state passes
through zero and vanishes. For values smaller than this critical 
value of $\lambda$ the integrand of the coupling-constant 
integral is not defined and therefore cannot be applied.
To cure this problem we introduce hard walls 
separated by a distance $L$ as a reference potential. 
Thus, all levels in the well stay bound and the integrand 
is defined for the entire range $\lambda \in [0,1]$. By taking 
the limit $L\to\infty$ the coupling-constant calculation converges 
to the case where only the original potential $v(x)$ is present. 
\begin{figure}[htb]
\begin{center}
\includegraphics[angle=0,width=8cm]{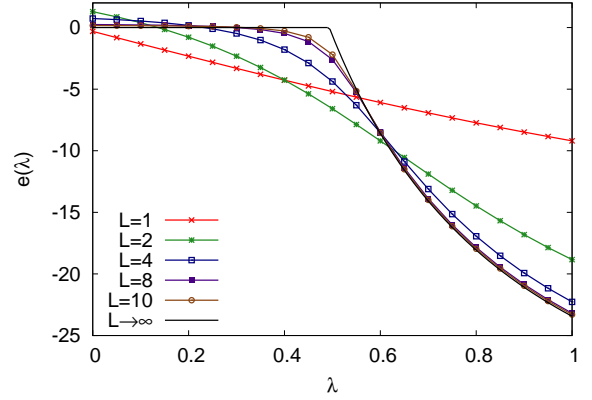}
\end{center}
\caption{\co 
Plot of $e(\lambda)$
for the second level as a function of the coupling-constant $\lambda$ 
and for different values of $L$ 
denoting the distance between the hard walls.}
\label{f:eg_j2}
\end{figure}
We demonstrate this explicitly by calculating the quantity 
\ben\label{elambda}
e(\lambda) = E_0(L) + \int_{-L/2}^{L/2} dx\ n[v^\lambda](x)\, v(x)
\een
for the second bound state and a well-depth of $40$. This 
is the $\lambda$-dependent integrand of \Eq{EHF} with 
$E_0(L) = 4\,\pi^2/(2L^2)$ denoting the reference energy of the
second bound state in an infinite square well; integrating $e(\lambda)$ 
over the coupling constant yields the orbital energy of the second bound state.
We plot $e(\lambda)$ for increasing $L$ in \Fig{f:eg_j2}:
the critical value of $\lambda$, below which the second level vanishes,
is indicated by the knee-like feature at $\lambda \approx 0.5$. 
The calculation in the simulation box converges to the exact result 
of the finite square well in a continuous manner as the separation 
between the walls is made larger.   This demonstrates how, in principle, 
the DSA can be applied for potentials that vanish at infinity, but also 
shows that the coupling-constant dependence may become quite strong.

\section{Definition of a \emph{ffunctional}}
\label{a:ffunctional}

Here we explain the meaning of the term \emph{ffunctional},
which first appears as $\mathcal{F}\cc_{n}[v]$ in \Eq{Fcc}, 
where we introduce the universal functional in terms of 
a coupling-constant expression.
Simply speaking, a \emph{ffunctional} maps a functional to a functional,
e.g., $\mathcal{F}\cc_{n}[v]$ takes the functional $n[w](\br)$ -- the density 
as a functional of some potential $w(\br)$ -- and creates a  
new functional of the external potential.

\begin{figure}[htb]
\begin{center}
\includegraphics[angle=0,width=7cm]{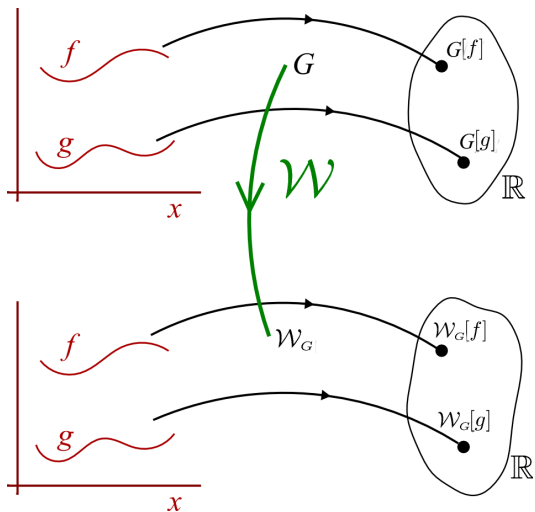}
\end{center}
\caption{\co Cartoon illustrating the concept of 
	a \emph{ffunctional}.}
\label{f:ffunctional}
\end{figure}

To understand this concept consider the cartoon in Fig.~\ref{f:ffunctional} 
and the following example: 
Assume a particular \emph{ffunctional} that is defined as 
\ben
\mathcal{W}_G[f] = \int_{-\infty}^\infty d^3r\ f^\alpha(\br)
\een
with $\alpha = \int_0^1 d\lambda\ G[\lambda q]$.
Now for a given functional $G[q]$, such as 
\ben
G[q] = \int_{-\infty}^\infty d^3r\ q^2(\br)\,,
\een
for which $\alpha=\int_{-\infty}^\infty d^3r\ q^2(\br)/3$
(assuming a well-behaved function $q(\br)$),
the \emph{ffunctional} $\mathcal{W}$ maps $G[q]$ to 
$\mathcal{W}_G[q]$. However, a different choice for $G[q]$
would have resulted in a different $\mathcal{W}_G$.

\bibliography{master_AC}
\label{page:end}
\end{document}